\documentclass[12pt,draftclsnofoot,journal,letterpaper,oneside,onecolumn]{IEEEtran}
\usepackage{mathrsfs}
\usepackage{array}
\usepackage{amssymb}
\usepackage{amsmath}
\usepackage{amsfonts}
\usepackage{graphicx}
\usepackage{subfigure}
\usepackage{cite}
\usepackage{enumerate}
\usepackage{url}

\usepackage{algorithm}
\usepackage{algorithmic}

\DeclareMathOperator*{\argmax}{argmax}

\begin{document}

\newtheorem{definition}{\bf Definition}
\newtheorem{proposition}{\bf Proposition}

\title{\Huge{Overlapping Coalition Formation Games for Emerging Communication Networks}}

\author{
\IEEEauthorblockN{
\small{Tianyu Wang}\IEEEauthorrefmark{1},
\small{Lingyang Song}\IEEEauthorrefmark{1},
\small{Zhu Han}\IEEEauthorrefmark{2},
\small{and Walid Saad}\IEEEauthorrefmark{3} \\}
\IEEEauthorblockA{
\IEEEauthorrefmark{1}\small{School of Electrical Engineering and Computer Science, Peking University, Beijing, China, \\ Email: tianyu.alex.wang@pku.edu.cn; lingyang.song@pku.edu.cn} \\
\IEEEauthorrefmark{2}\small{Electrical and Computer Engineering Department, Computer Science Department, University of Houston, Houston, TX, USA, \\ Email: zhan2@uh.edu}\\
\IEEEauthorrefmark{3}\small{Wireless@VT, Bradley Department of Electrical and Computer Engineering, Virginia Tech, Blacksburg, VA, USA, Email: walids@vt.edu} \\}
\thanks{This research was supported by the U.S. National Science Foundation under Grants CNS-1460316 and AST-1506297, NSF CCF-1456921, Grant CNS-1443917, Grant ECCS-1405121, and Grant NSFC 61428101.}
}

\maketitle

\begin{abstract}

Modern cellular networks are witnessing an unprecedented evolution from classical, centralized and homogenous architectures into a mix of various technologies, in which the network devices are densely and randomly deployed in a decentralized and heterogenous architecture. This shift in network architecture requires network devices to become more autonomous and, potentially, cooperate with one another. Such cooperation can, for example, take place between interfering small access points that seek to coordinate their radio resource allocation, nearby single-antenna users that can cooperatively perform virtual MIMO communications, or even unlicensed users that wish to cooperatively sense the spectrum of the licensed users. Such cooperative mechanisms involve the simultaneous sharing and distribution of resources among a number of overlapping cooperative groups or coalitions. In this paper, a novel mathematical framework from cooperative games, dubbed \emph{overlapping coalition formation games} (OCF games), is introduced to model and solve such cooperative scenarios. First, the concepts of OCF games are presented, and then, several algorithmic aspects are studied for two main classes of OCF games. Subsequently, two example applications, namely, interference management and cooperative spectrum sensing, are discussed in detail to show how the proposed models and algorithms can be used in the future scenarios of wireless systems. Finally, we conclude by providing an overview on future directions and applications of OCF games.

\end{abstract}

\newpage

\section{Introduction}%

Recent studies show that, driven by emergence of highly capable devices such as smartphones and resource demanding wireless services such as video streaming, the demand for wireless capacity will increase roughly 1000x, compared to the current 4G networks, by the year 2020~\cite{Cisco-2014}. In order to overcome this wireless capacity crunch, an evolutionary architecture has been introduced in the next-generation 5G networks, in which low-cost and lower-power small cell base stations (SBSs) are densely and randomly deployed within the coverage of macrocell base stations (MBSs)~\cite{ABCHLSZ-2014} and~\cite{A-2013}. This new architecture has shown its great potential to improve the capacity and coverage of wireless cellular systems. However, such extreme network densification makes it challenging to manage the cellular system at various levels that include interference control, coverage optimization, load balancing, mobility robustness optimization, and energy management. Many of the existing solutions require autonomous cooperation between network devices. For example, SBSs can cooperate for performing coordinated multipoint (CoMP) transmissions~\cite{OHI-2012} or for coordinating their interference via techniques such as interference alignment~\cite{PBSDL-2013}. Similarly, user cooperation can take place in order to further exploit user diversity, e.g., user devices can relay for each other by using device-to-device~(D2D) communications~\cite{KLNK-2014}, or form cooperative groups to support virtual multiple-input multiple-output (MIMO)~\cite{FWGLZ-2013} transmissions.

Despite in emerging wireless networks, most existing optimization techniques are restricted to addressing the cooperation problem for centralized and homogenous wireless systems in which cooperation is a privilege rather than a necessity. In order to provide a more flexible framework for the cooperative behaviors of future wireless systems, each device in the network can be treated as an individual decision maker that act on its own principles, which naturally leads to game-theoretic approaches where multiple players form a stable and efficient network operating point in a self-organizing manner~\cite{HNSBH-2011}. In particular, the framework of \emph{cooperative game theory} provides the necessary tools for modeling and developing self-organizing techniques for forming cooperative groups or coalitions between network devices, based on the mutual benefit and costs for cooperation. Indeed, cooperative games, in general, and coalition formation games (CF games), in particular, have become a popular tool for analyzing wireless networks~\cite{SHDHB-2009,SHDH-2008,SHDHB-2009b,SHBDH-2009,HP-2009,SHBDH-2010,SHHNH-2011,PBSVL-2011,WCZW-2012,WSH-2013}.

In~\cite{SHDH-2008}, single antenna users self-organize into multiple coalitions and share their antennas in each coalition to form a virtual MIMO system, and hence benefit from spatial diversity or multiplexing. In~\cite{SHDHB-2009b}, secondary users (SUs) form multiple coalitions and combine their individual sensing data at the coalition head to perform cooperative spectrum sensing, and hence improve their sensing performance. In~\cite{SHBDH-2009}, wireless transmitters form coalitions to coordinate their transmissions so as to improve their physical layer security. In~\cite{HP-2009}, coalition games have been used to overcome the curse of boundary nodes, in which boundary nodes use cooperative transmissions to help the backbone nodes in the middle of the network. In~\cite{SHBDH-2010}, hedonic coalition formation games are utilized to model the task allocations problem, in which a number of wireless agents are required to collect data from several arbitrarily located tasks using wireless transmissions. In~\cite{SHHNH-2011}, a coalition formation scheme is proposed for road-side units in vehicular networks to improve the diversity of the information circulating in the network while exploiting the underlying content-sharing vehicle-to-vehicle communication network. In~\cite{PBSVL-2011}, a cooperative model based on coalition formation game is proposed to enable femtocells to improve their performance by sharing spectral resources, minimizing the number of collisions, and maximizing the spatial reuse. In~\cite{WCZW-2012}, coalition formation games are utilized to strike a balance between the QoS provisioning and the energy efficiency in a clustered wireless sensor network. However, most of this existing body of work focuses on coalition formation models in which the players form separate coalitions and get payoffs from the single coalition they join. In future wireless systems, communication nodes are equipped with more powerful devices that are able to utilize multiple resources in a more flexible way. This makes it possible and necessary to allow nodes to participate in multiple \emph{overlapping coalitions}, and, subsequently, receive payoffs from all coalitions they participate~\cite{ZSHS-2014, WSHS-2013, DWSH-2013, BSSVDC-2013}. For example, a multiple antenna user may join multiple coalitions by devoting its antennas into different groups of users, and benefit from multiple virtual MIMO transmissions, and also, an SBS with multiple neighboring SBSs needs to coordinate its transmissions with multiple groups of SBSs, so as to avoid inner-channel interferences from all its neighbors. In these scenarios, the coalitions formed by communication nodes are overlapping, and each node receives payoffs from the multiple coalitions it joins. Even though there are many other available modeling tools for analyzing cooperation in wireless networks~\cite{CGYZ-2014}, we focus on the methods that belong to cooperative game theory.

The main contribution of this paper is to present an introduction to a novel mathematical framework from cooperative games, \emph{overlapping coalition formation games}~(OCF games), which provides the necessary analytical tools for analyzing how players in a wireless network can cooperate by joining, simultaneously, multiple overlapping coalitions~\cite{CEMJ-2010,ZE-2011,ZCE-2012}. First, in Section~\uppercase\expandafter{\romannumeral2}, we introduce the basic concepts of OCF games in general, and develop two polynomial algorithms for two classes of OCF games. Then, in Sections~\uppercase\expandafter{\romannumeral3} and~\uppercase\expandafter{\romannumeral4}, based on~\cite{ZSHS-2014} and~\cite{WSHS-2013}, we present two emerging applications of OCF games in small cell-based heterogeneous networks (HetNets) and cognitive radio networks, in order to show the advantages of forming overlapping coalitions compared with the traditional non-overlapping coalitional games. In Section~\uppercase\expandafter{\romannumeral5}, we conclude by summarizing the potential applications of OCF games in future wireless networks.

\section{Cooperative Games with Overlapping Coalitions}%

In this section, we formally introduce the notion of cooperative games with overlapping coalitions, or OCF games. In Section~\uppercase\expandafter{\romannumeral2}.~A, we present the basic model of OCF games and illustrate the overlapping gain compared to traditional cooperative games with non-overlapping coalitions. In Section~\uppercase\expandafter{\romannumeral2}.~B, we focus on one key stability notion in OCF games, $\mathcal{A}$-core, which is a direct extension of the core from traditional cooperative games. In this regard, we show that the computation of ``stable" outcomes in the sense of the $\mathcal{A}$-core can generally be intractable. Therefore, in Section~\uppercase\expandafter{\romannumeral2}.~C, we identify several constraints that lead to tractable subclasses of OCF games, and provide efficient algorithms for solving games that fall under these subclasses.

\subsection{Basic Concepts}%

Game theory is a mathematical tool that analyzes systems with multiple decision makers having interdependent objectives and actions. The decision makers, which are usually referred to as \emph{players}, will interact and obtain individual profits from the resulting outcome. In cooperative games, the players can form cooperative groups, or \emph{coalitions}, to jointly increase their profits in a game. In traditional cooperative games, the players are typically assumed to form disjoint, non-overlapping coalitions, and they only cooperate with players within the same coalition. However, there are situations in which some players may be involved in multiple coalitions simultaneously. In such cases, these players may need to split their resources among the coalitions that they participate in. For example, a multi-mode wireless terminal may access base stations from different networks and it needs to distribute its traffic load among these networks~\cite{BSSVDC-2013}. In such situations, some coalitions (cells of different systems) may involve some of the same players (multi-mode terminals), and therefore may overlap with one another. Now, we formally introduce the mathematical tool to model these ``overlapping" situations, \emph{cooperative games with overlapping coalitions}, or \emph{OCF games}.

In OCF games, each player possesses a certain amount of resources such as time, power or money. In order to obtain individual profits, the players form coalitions by contributing a portion of their resources and receive payoffs from the devoted coalitions. A \emph{coalition} can be represented by the resource vector contributed by its coalition members, i.e., $\boldsymbol{r} = (r_1,r_2,\ldots,r_N)$, where $0 < r_i < R$ represents player $i$'s resources that are contributed to this coalition. For each coalition $\boldsymbol{r}$, the \emph{coalition value} is decided by a function $v: [0,R]^N \rightarrow \mathbb{R}^+$, which represents the total payoff that the players can get from a cooperative coalition. The coalition value can be divided to the coalition members based on specific rules, e.g. the value can be equally divided among coalition members, or it can be divided based on the contribution of each member. We denote by $\boldsymbol{x}$ as the payoff allocation rule, and accordingly, the individual payoff that player $i$ receives from coalition $\boldsymbol{r}$ is denoted by $x_i(\boldsymbol{r})$. The players may decide to devote different amount of resources into different coalitions, so as to maximize its individual payoff $p_i(\boldsymbol{\pi},\boldsymbol{x}) = \sum\nolimits_{\boldsymbol{r} \in \boldsymbol{\pi}} x_i(\boldsymbol{r})$, where $\boldsymbol{\pi}$ represents the set of all coalitions $\boldsymbol{\pi} = \{\mathbf{r}^1, \mathbf{r}^2, \ldots, \mathbf{r}^K\}$ formed by the players. Note that these coalitions may have common members, and thus, they form an \emph{overlapping coalition structure}. Compared with the traditional CF games, OCF games allow the players to form an overlapping coalition structure and get payoffs from multiple coalitions. Therefore, the overlapping structure may provide more flexibility for the players to utilize their resources, which enables the coalitions to be better organized and potentially leads to outcomes with higher payoffs. The following example clearly shows the potential advantage of OCF games.

\begin{figure}
\centering
\subfigure[the CF game model]{
\label{e1} 
\includegraphics[width=2.8in]{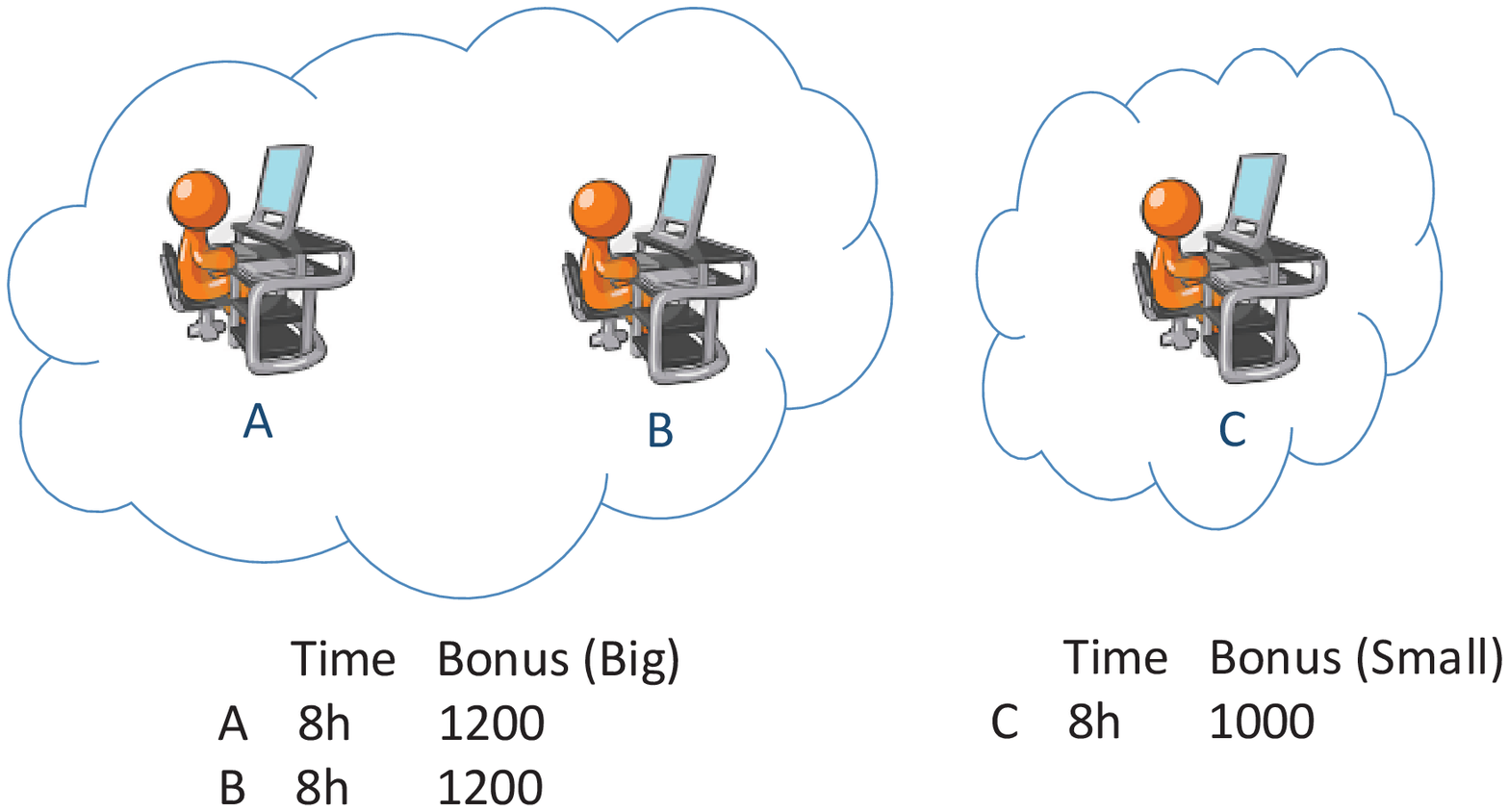}}
\hspace{0.3in}
\subfigure[the OCF game model]{
\label{e2} 
\includegraphics[width=2.8in]{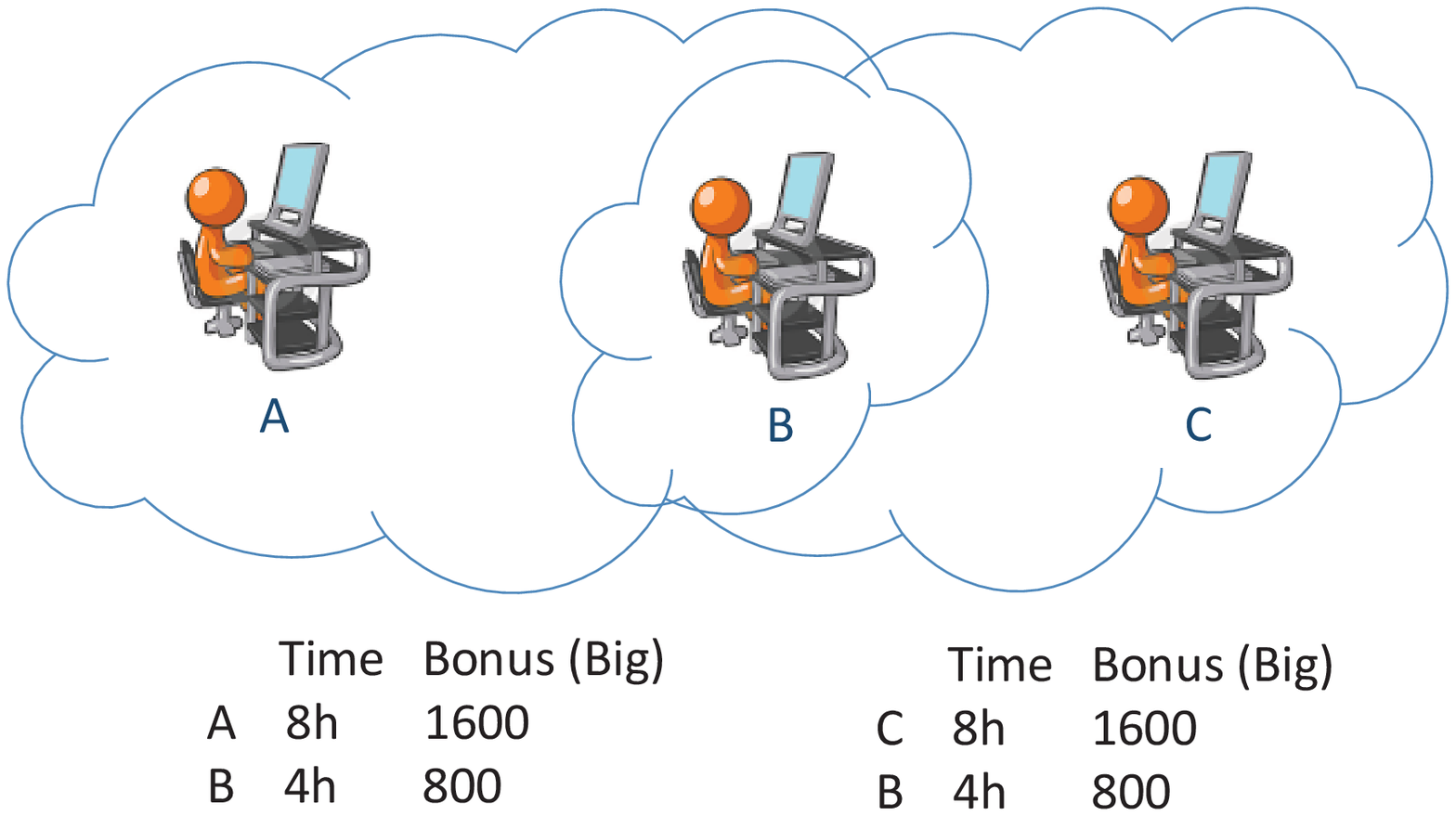}}
\caption{An example to show the differences between OCF games and traditional CF games.}
\label{example} 
\end{figure}

\textbf{Example:} Consider a software company with three developers $A, B$ and $C$. Each developer works $8$ hours a day. There are two types of projects in the company, big projects and small projects. A big project requires $12$ man-hours per day and provides a $2400$ bonus, and a small project requires $8$ man-hours per day and provides a $1000$ bonus. We assume the bonus is divided to the participating developers according to their devoted time. In a traditional CF game, the players can only form disjoint coalitions, and the optimal coalition structure is $\{\{A,B\},\{C\}\}$, as seen in Fig.~\ref{e1}, i.e., developers $A$ and $B$ work together to accomplish a big project and developer $C$ work alone for a small project. The total payoffs of $A, B$ and $C$ are then given by $(1200,1200,1000)$. In an OCF game, the players can split their time into different coalitions, and the optimal coalition structure is $\{(8,4,0),(0,4,8)\}$, as seen in Fig.~\ref{e2}, i.e., developers $A$ and $B$ devote $8$ and $4$ hours to accomplish a big project, and developers B and C devote $4$ and $8$ hours to accomplish another big project. The payoffs of $A, B$ and $C$ are then given by $(1600,1600,1600)$.

\subsection{Stability Notions}%

In cooperative games, one must seek a stable outcome, i.e., a coalition structure in which no set of players can deviate and obtain a new structure that increases all their payoffs~\cite{SHDHB-2009,ZCE-2012}. In traditional CF games, the deviating players, or deviators, leave their original coalitions and form a new coalition, the value of which is the total payoff that the deviators can get from their deviation. If there exists a payoff division that makes all the deviators achieve a higher payoff compared with the original coalition structure, we say the deviation is profitable. If there exists no profitable deviation, we say the structure is stable. Particularly, if the players have an incentive to form the grand coalition that includes all players, the cooperative game is called a canonical game and the set of all stable payoff divisions, corresponding the grand coalition, is called the core of such a game.

Compared with CF games, defining notions for stability in OCF games is more complicated due to the overlapping property. In OCF games, instead of forming a single coalition, the deviators may form multiple coalitions that overlap one another thus complicating the computation of the maximal total payoff of such an overlapping coalition structure. Also, the deviators in OCF games may partially deviate from the original coalitions by withdrawing a portion of their
resources while maintaining their other resources in their original coalitions. Therefore, one must precisely define how the non-deviators in the original coalitions will react to such a deviation and how much payoff the deviators can get from those partially deviated coalitions. To this end, we now present the \emph{$\mathcal{A}$-core}, a stability notion suitable for OCF games, which is an extension of the core notion from traditional CF games.

We begin by defining a deviation in OCF games. Given a coalition structure $\boldsymbol{\pi}$ and a set of players $\mathcal{S}$ that attempts to deviate the structure, the coalitions in $\boldsymbol{\pi}$ can be divided into two groups: the coalitions that only involve players in $\mathcal{S}$, denoted by $\boldsymbol{\pi}|_\mathcal{S}$, and the coalitions that involves players other than $\mathcal{S}$, given by $\boldsymbol{\pi} \backslash \boldsymbol{\pi}|_\mathcal{S}$. Since coalitions $\boldsymbol{\pi}|_\mathcal{S}$ are fully controlled by the deviators, they should be seen as pure resources withdrawn by the deviators. While coalitions $\boldsymbol{\pi} \backslash \boldsymbol{\pi}|_\mathcal{S}$, which include both deviators and non-deviators, should be considered as coalitions that are partially deviated by the deviators. We define $D(\boldsymbol{\pi}) = \{\boldsymbol{d}(\boldsymbol{r}) | \boldsymbol{r} \in \boldsymbol{\pi} \backslash \boldsymbol{\pi}|_\mathcal{S} \}$ as the resources withdrawn from $\boldsymbol{\pi} \backslash \boldsymbol{\pi}|_\mathcal{S}$, where $\boldsymbol{d}(\boldsymbol{r})$ is the resources that $\mathcal{S}$ withdraw from coalition $\boldsymbol{r}$.

The deviators $\mathcal{S}$ can form an overlapping coalition structure using both the withdrawn resources $D(\boldsymbol{\pi})$, and the resources of their own $\boldsymbol{\pi}|_\mathcal{S}$. We denote by $\boldsymbol{W}_\mathcal{S}$ as the sum available resources of the deviators $\mathcal{S}$, and $\boldsymbol{\Pi}(\boldsymbol{W}_\mathcal{S})$ as the set of all possible coalition structures that can be formed using $\boldsymbol{W}_\mathcal{S}$. The optimal coalition structure formed by $\mathcal{S}$ is then given by $\boldsymbol{\pi}(\boldsymbol{W}_\mathcal{S}) = \argmax\nolimits_{\boldsymbol{\pi} \in \boldsymbol{\Pi}(\boldsymbol{W}_\mathcal{S})} \left\{ \sum\nolimits_{\boldsymbol{r} \in \boldsymbol{\pi}} v(\boldsymbol{r}) \right\}$. Note that the deviators may also receive payoffs from the coalitions that they partially deviate, i.e., $\boldsymbol{\pi} \backslash \boldsymbol{\pi}|_\mathcal{S}$. We formally define the \emph{arbitration function} $\mathcal{A}_{\boldsymbol{r}} ( \boldsymbol{\pi}, \boldsymbol{x}, D(\boldsymbol{\pi}), \mathcal{S} )$, which represents the total payoff that the deviators $\mathcal{S}$ will receive from coalition $\boldsymbol{r} \in \boldsymbol{\pi} \backslash \boldsymbol{\pi}|_\mathcal{S}$.

\begin{definition}\label{A-core}
A deviation on a player set $\mathcal{S}$ is said to be \emph{$\mathcal{A}$-profitable} if and only if
\begin{equation} \label{A-profitable}
\sum\limits_{i \in \mathcal{S}} p_i(\boldsymbol{\pi},\boldsymbol{x}) < \sum\limits_{\boldsymbol{r} \in \boldsymbol{\pi}(\boldsymbol{W}_\mathcal{S})} v(\boldsymbol{r}) + \sum\limits_{\boldsymbol{r} \in \boldsymbol{\pi} \backslash \boldsymbol{\pi}|_\mathcal{S}} \mathcal{A}_{\boldsymbol{r}} (\boldsymbol{\pi}, \boldsymbol{x}, D(\boldsymbol{\pi}), \mathcal{S}).
\end{equation}
If there exists no $\mathcal{A}$-profitable deviation for any player set, then the coalition structure $\boldsymbol{\pi}$ is said to be in the \emph{$\mathcal{A}$-core}, or \emph{$\mathcal{A}$-stable}.
\end{definition}

The coalition structures in $\mathcal{A}$-core represent the stable structures in which no set of players have the motivation to deviate from the current structure. We note that the definition of $\mathcal{A}$-core depends on the specific form of the arbitration function. According to different assumptions on players and different applicable scenarios \cite{CEMJ-2010,ZCE-2012}, three of the mostly used arbitration functions are described as follows:
\begin{enumerate}
    \item \emph{c-core}: if the players are very conservative in cooperating with deviators, the deviators may receive no payoffs from any original coalitions in $\boldsymbol{\pi} \backslash \boldsymbol{\pi}|_\mathcal{S}$, even if they still contribute to these coalitions, i.e., $\mathcal{A}_{\boldsymbol{r}} (\boldsymbol{\pi}, \boldsymbol{x}, D(\boldsymbol{\pi}), \mathcal{S}) \equiv 0$. This is called the \emph{conservative arbitration function}, and the stability notion is referred to as \emph{c-core}.
    \item \emph{r-core}: if the players are more lenient, the deviators can still get their original payoffs from the coalitions that are not influenced by their deviation, i.e., $\mathcal{A}_{\boldsymbol{r}} (\boldsymbol{\pi}, \boldsymbol{x}, D(\boldsymbol{\pi}), \mathcal{S}) = \sum\nolimits_{i \in \mathcal{S}} x_i(\boldsymbol{r})$ for all coalition $\boldsymbol{r}$ with $\boldsymbol{d}(\boldsymbol{r}) = 0$. This is called the \emph{refined arbitration function}, and the stability notion is referred to as \emph{r-core}.
    \item \emph{o-core}: the players can be highly generous that they allow the deviators to keep all the ``leftover" payoff as long as the non-deviators' original payoffs are ensured to be unchanged, i.e., $\mathcal{A}_{\boldsymbol{r}} (\boldsymbol{\pi}, \boldsymbol{x}, D(\boldsymbol{\pi}), \mathcal{S}) = v(\boldsymbol{r} - \boldsymbol{d}(\boldsymbol{r})) - \sum\nolimits_{i \in \mathcal{N} \backslash \mathcal{S}} x_i(\boldsymbol{r})$. This is called the \emph{optimistic arbitration function}, and the stability notion is referred to as \emph{o-core}.
\end{enumerate}
Hereinafter, we adopt the \emph{o-core} as the stability notion due to its computational advantage that we will explain in the next subsection.

\subsection{Algorithms for o-Stable Outcomes}%

To avoid the difficulty in representing the value function $v$ and the arbitration function $\mathcal{A}$, we assume that the resources are given by integers such that $R \in \mathbb{Z}^+$ and the players can only divide their resources in a discrete manner. Such games are referred to as \emph{discrete OCF games} and they apply to practical systems. Also, due to the cost of information exchange between deviators, we assume the number of deviators in a deviation is preliminarily bounded. We denote by $S$ as the upper bound of deviation size, i.e., $|\mathcal{S}| \le S$ for any deviation on any player set $\mathcal{S}$. It has been shown that computing an $\mathcal{A}$-stable outcome of an OCF game is generally a challenging problem~\cite{ZCE-2012}. However, if we only consider the o-core notion and identify several constraints on the game, there exists efficient algorithms that lead to o-stable outcomes of such games.

\begin{proposition}\label{o-property}
Any o-profitable deviation on any player set $\mathcal{S} \subseteq \mathcal{N}$ will increase the \emph{social welfare}, which is defined as the total value of all coalitions in the outcome, i.e., $\sum\nolimits_{\boldsymbol{r} \in  \boldsymbol{\pi}} v(\boldsymbol{r})$.
\end{proposition}

\begin{IEEEproof}
The social welfare can be written as $SW = \sum\nolimits_{\boldsymbol{r} \in  \boldsymbol{\pi} \backslash \boldsymbol{\pi}|_{\mathcal{S}}} v(\boldsymbol{r}) + \sum\nolimits_{\boldsymbol{r} \in \boldsymbol{\pi}|_{\mathcal{S}}} v(\boldsymbol{r})$, where the first item represents the sum value of $\boldsymbol{\pi} \backslash \boldsymbol{\pi}|_{\mathcal{S}}$, and the second item represents the sum value of $\boldsymbol{\pi}|_{\mathcal{S}}$. After the deviation of $\mathcal{S}$, the deviators withdraw $\boldsymbol{d}(\boldsymbol{r})$ resources from each coalition $\boldsymbol{r} \in \boldsymbol{\pi} \backslash \boldsymbol{\pi}|_{\mathcal{S}}$, and the coalition is reduced to $\boldsymbol{r} - \boldsymbol{d}(\boldsymbol{r})$. Meanwhile, the deviators form an optimal coalition structure $\boldsymbol{\pi}(\boldsymbol{W}_\mathcal{S})$. Therefore, the social welfare after deviation is given by $SW' = \sum\nolimits_{\boldsymbol{r} \in  \boldsymbol{\pi} \backslash \boldsymbol{\pi}|_{\mathcal{S}}} v(\boldsymbol{r} - \boldsymbol{d}(\boldsymbol{r})) + \sum\nolimits_{\boldsymbol{r} \in \boldsymbol{\pi}(\boldsymbol{W}_\mathcal{S})} v(\boldsymbol{r})$. According to the definition of o-core, we can easily prove that the social welfare is increased by the deviation on $\mathcal{S}$, i.e., $SW'>SW$. Therefore, we conclude that any o-profitable deviation will increase the social welfare.
\end{IEEEproof}

In most practical problems, the social welfare is bounded due to the limited resources. For such games, Proposition \ref{o-property} implies that the game must converge to an o-stable outcome after finite o-profitable deviations. Therefore, we can compute an o-stable outcome by iteratively computing an o-profitable deviation to the current outcome. However, finding an o-profitable deviation is a challenging problem. First, without further restrictions on deviations, the number of potential deviations can be extremely large. Second, deciding whether a deviation is o-profitable requires solving the optimal coalition structure problem, which is not a straightforward problem. Now, we define two subclasses of OCF games, namely $K$-coalition OCF games and $K$-task OCF games, and we provide efficient algorithms to compute o-stable outcomes in such games, respectively.

In a $K$-coalition OCF game, we assume each player can contribute to at most $K$ coalitions. This assumption is reasonable in many practical systems due to geographical constraints, communication cost, or lack of information. For example, a mobile user can only connect to limited base stations that are close to it. Therefore, for a group of deviators $\mathcal{S}$ in a $K$-coalition game, the number of possible deviations is bounded by $(R+1)^{SK}$. Since there are $C_N^S$ groups of possible deviators, the total number of possible deviations is then bounded by $C_N^S (R+1)^{SK} = \mathcal{O}(N^S)$, which is polynomial in $N$. For any deviation on $\mathcal{S}$, the deviators need to calculate the optimal coalition structure $\boldsymbol{\pi}(\boldsymbol{W}_\mathcal{S})$ to decide whether the deviation is o-profitable. We define the \emph{superadditive cover} of $v$ to be the function $v^*: [0,R]^N \rightarrow \mathbb{R}^+$, such that $v^*(\boldsymbol{W}) = \max\nolimits_{\boldsymbol{\pi} \in \mathcal{\boldsymbol{\Pi}}(\boldsymbol{W})} \left\{ \sum\nolimits_{\boldsymbol{r} \in \boldsymbol{\pi}} v(\boldsymbol{r}) \right\}$ for any resource vector $\boldsymbol{W}$. Briefly, $v^*(\boldsymbol{W})$ is the maximal total value that the players can generate by forming overlapping coalitions when their total resources are given by $\boldsymbol{W}$. We observe that $v^*(\boldsymbol{W}) = \max \left\{ v^*(\boldsymbol{W} - \boldsymbol{r}) + v(\boldsymbol{r}) | \boldsymbol{r} \preccurlyeq \boldsymbol{W} \right\}$, which is a recurrence relation for a discrete-time dynamic system. Thus, we can use the dynamic programming algorithm to calculate $v^*(\boldsymbol{W})$. Given the values of $v^*(\boldsymbol{W}')$ for all $\boldsymbol{W}' \preccurlyeq \boldsymbol{W}$, the computation of $v^*(\boldsymbol{W})$ requires $(R+1)^S$ times of computing $v$. Therefore, the entire computation of $v^*(\boldsymbol{W})$ requires at most $(R+1)^S (R+1)^S = (R+1)^{2S}$ times of computing $v$. When $v^*(\boldsymbol{W})$ is calculated, we can trace backward the optimal path and achieve every coalition in the optimal coalition structure $\boldsymbol{\pi}(\boldsymbol{W})$. Therefore, the optimal coalition structure $\boldsymbol{\pi}(\boldsymbol{W}_\mathcal{S})$ can be calculated in time $(R+1)^{2S}$. Therefore, we can calculate an o-profitable deviation in time $C_N^S (R+1)^{SK} (R+1)^{2S} = C_N^S (R+1)^{S(K+2)} = \mathcal{O}(N^S)$, which is polynomial in $N$. The algorithm for $K$-coalition games is shown in Table~\ref{K-coalition}.

\begin{table}[!t]
\renewcommand{\arraystretch}{2.0}
\caption{Algorithm for $K$-coalition games}
\label{K-coalition}
\centering
\begin{tabular}{p{160mm}}

\hline

Input an initial outcome $(\boldsymbol{\pi}_0,\boldsymbol{x}_0)$.

\begin{algorithmic} [1]
\STATE $\boldsymbol{\pi} \leftarrow \boldsymbol{\pi}_0$ $\%$ initial coalition structure
\STATE $\boldsymbol{x} \leftarrow \boldsymbol{x}_0$ $\%$ initial payoffs
\WHILE{there exists an o-profitable deviation on a player set $\mathcal{S}$}
\STATE $(\boldsymbol{r} - \boldsymbol{d}(\boldsymbol{r})) \leftarrow \boldsymbol{r}$ for all $\boldsymbol{r} \in \boldsymbol{\pi} \backslash \boldsymbol{\pi}|_\mathcal{S}$
\STATE $\boldsymbol{\pi}( \boldsymbol{W}_\mathcal{S} ) \leftarrow \boldsymbol{\pi}|_\mathcal{S}$
\STATE decide new payoffs $\boldsymbol{x}$
\ENDWHILE
\STATE $\mathcal{\boldsymbol{\pi}}_f \leftarrow \mathcal{\boldsymbol{\pi}}$ $\%$ final coalitional structure
\STATE $\boldsymbol{x}_f \leftarrow \boldsymbol{x}$ $\%$ final payoffs
\end{algorithmic}

Output an o-stable outcome $(\boldsymbol{\pi}_f,\boldsymbol{x}_f)$. \\

\hline

\end{tabular}
\end{table}

In a $K$-task OCF game, each coalition in the game corresponds to a specific task and each player can only contribute to $K$ tasks. Being different from $K$-coalition OCF games, the number of coalitions in a $K$-task OCF game is strictly limited by the number of tasks, which are predetermined by the considered problem. For example, in a software company, the available projects are predetermined and the developers cannot form coalitions to generate new projects but only divide his time among the existing ones. Since the number of coalitions is fixed in a $K$-task OCF game, a deviation will not form new coalition structures but only move resources among the existing coalitions, and thus, we refer to deviation as \emph{transfer} in $K$-task OCF games. The number of possible transfers is now given by $C_N^S [K^2(R+1)]^S = \mathcal{O}(N^S)$, which is polynomial in $N$. Since the deviators do not form an overlapping coalition structure, their payoffs can be easily calculated using the arbitration function. Therefore, an o-profitable deviation of a $K$-task OCF game can be calculated in time $\mathcal{O}(N^S)$. The algorithm for $K$-task games is shown in Table~\ref{K-task}.

\begin{table}[!t]
\renewcommand{\arraystretch}{2.0}
\caption{Algorithm for $K$-task games}
\label{K-task}
\centering
\begin{tabular}{p{160mm}}

\hline

Input an initial outcome $(\boldsymbol{\pi}_0,\boldsymbol{x}_0)$.

\begin{algorithmic} [1]
\STATE $\boldsymbol{\pi} \leftarrow \boldsymbol{\pi}_0$ $\%$ initial coalition structure
\STATE $\boldsymbol{x} \leftarrow \boldsymbol{x}_0$ $\%$ initial payoffs
\WHILE{there exists an o-profitable transfer on a player set $\mathcal{S}$}
\STATE $(\boldsymbol{r} - \boldsymbol{d}(\boldsymbol{r})) \leftarrow \boldsymbol{r}$ for all $\boldsymbol{r} \in \boldsymbol{\pi}$
\STATE decide new payoffs $\boldsymbol{x}$
\ENDWHILE
\STATE $\mathcal{\boldsymbol{\pi}}_f \leftarrow \mathcal{\boldsymbol{\pi}}$ $\%$ final coalitional structure
\STATE $\boldsymbol{x}_f \leftarrow \boldsymbol{x}$ $\%$ final payoffs
\end{algorithmic}

Output an o-stable outcome $(\boldsymbol{\pi}_f,\boldsymbol{x}_f)$. \\

\hline

\end{tabular}
\end{table}

Given the polynomial algorithms for $K$-coalition games and $K$-task games, we then provide two example applications to show how the concepts and algorithms of OCF games can be utilized in wireless networks. Note that we restrict our model to single-resource scenarios in which the players only have one type of resources. However, this model can be extended to the multi-resource setting, by using a vector rather than a scalar to describe the contribution of a player, and all the concepts and algorithms can also be extended to such a case.

\section{Application Scenario~\uppercase\expandafter{\romannumeral1}: Interference Management in HetNets}%

\begin{figure}[!t]
\centering
\includegraphics[width=4.2in]{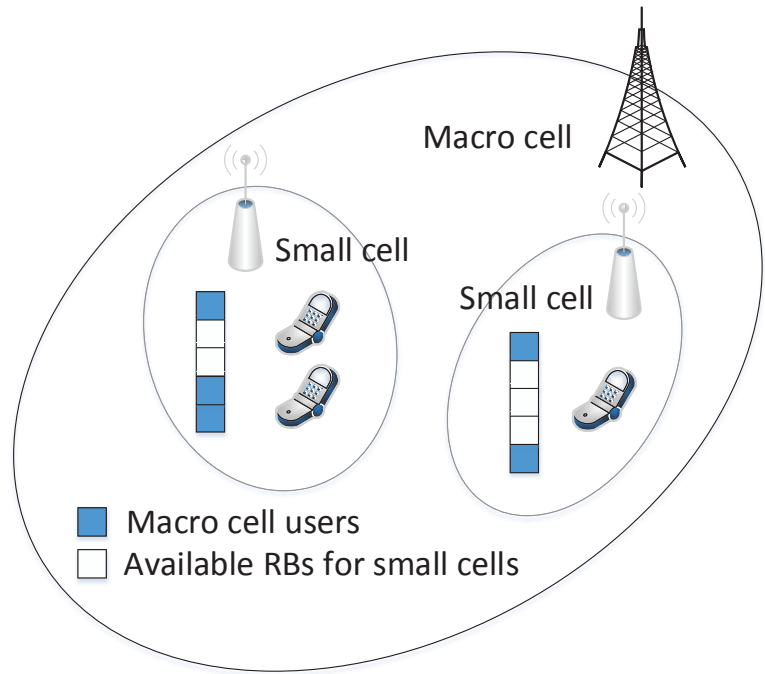}
\caption{Application scenario ~\uppercase\expandafter{\romannumeral1}: radio resource allocation in HetNets} \label{app1}
\end{figure}

In small cell-based HetNets, a large number of small cells may be randomly deployed in the same spectrum as the existing, macro-cellular network. Due to the large amount of small cells and their ad hoc nature of deployment, interference management is always one of the key challenges in HetNets~\cite{ZANM-2013}. There are many interference management techniques, such as successive interference cancellation (SIC), parallel interference cancellation (PIC) and multiuser detection (MUD)~\cite{Andrews-2005}. However, these techniques require global knowledge of the characteristics of the interfering channels, which generates a huge amount of backhaul traffic for small cells and makes it impractical when the number of small cells is large. In this case, distributed schemes become important, where no central controller is involved in the system and minimum information exchange is required among the small base stations~\cite{SS-2011}.

Game theory, due to its self-organizing characteristic, has also been widely utilized to design distributed approaches for interference management~\cite{HT-2010}. However, most of these approaches are based on non-cooperative games where no information exchange is allowed among small cells. In this paper, we propose a cooperative approach based on OCF games. We consider the downlink scenario in which the macro users are interfered by nearby SBSs, and small cell users are interfered by MBSs as well as nearby SBSs. To avoid interference between the macro network and small cells, the underlayed MBS can inform each SBS in coverage of the available RBs in the current slot, which is determined by the radio assignment of the macro users. However, interference between small cells still exists due to the lack of coordination between SBSs. This scenario is illustrated in Fig.~\ref{app1}. We will study how OCF games can be used to coordinate the interference between SBSs and improve the entire network performance~\cite{ZSHS-2014}.

\begin{figure}[!t]
\centering
\includegraphics[width=4.2in]{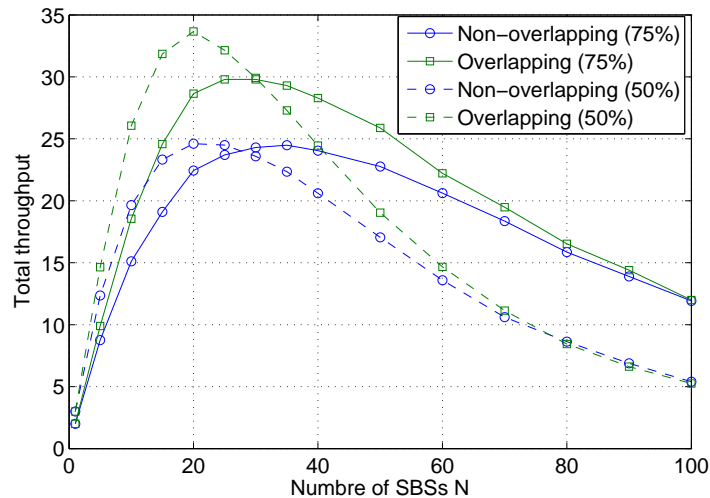}
\caption{The total throughput as a function of the number of SBSs in a $400 \times 400$ $m^2$ area with different levels of traffic load. The interference radius of each SBS is set as $100$ m.} \label{radio}
\end{figure}

For each SBS, the available RBs, which are decided by the MBS that covers it, are the resources that can be used to cooperate with other SBSs. For each available RB, the SBS can decide to leave it unoccupied so as to reduce the inter-cell interference, or to utilize it for downlink transmissions so as to improve the throughput. When the SBSs cooperate with each other by contributing a part of their available RBs, they form a coalition in which all contributed RBs are evenly distributed to the involved SBSs. To avoid interference inside the coalition, each RB can only be distributed to one SBS, and the SBSs can only utilize the RBs distributed to them. The value of a coalition is given by the total downlink throughput generated by the resources of this coalition. Note that the interference outside the coalition still exists and it should be considered in the calculation of coalition value. The value is then distributed to the involved SBSs as their payoffs when they actually use them in the downlink transmission. In order maximize their individual payoffs, the SBSs may form different overlapping coalitions by deviating from the current overlapping coalition structure. The dynamics can be seen as an OCF-game, which converges to an o-stable structure as we explained. Since the coalitions can only be formed by neighboring SBSs, the number of coalitions that a SBS can participate is limited. Therefore, the studied OCF game is a $K$-coalition OCF game. We use the developed algorithm in Table~\ref{K-coalition} and the performance is shown in Fig.~\ref{radio}.

In Fig.~\ref{radio}, we compare the developed algorithm with the situation of no overlapping in networks with different levels of traffic load. The values $50\%$ and $75\%$ represent the average rate between the number of required RBs by small cell users and the total available RBs for SBSs. When the SBSs are sparsely deployed, the SBSs seldom interfere each other, and the performance improvement by the developed algorithm is limited. As more SBSs are deployed in the area, the interference coordination becomes crucial and the developed algorithm improves the network performance by $20\%$ to $40\%$. When the SBSs are extremely dense, the inevitable interference between the coalitions dominates the network performance, and the advantage of the developed algorithm converges to zero. Also, when the traffic load is heavier, interference coordination can bring more benefits to the network, and thus, the developed algorithm performs better.

\section{Application Scenario~\uppercase\expandafter{\romannumeral2}: Cooperative Sensing}%

\begin{figure}[!t]
\centering
\includegraphics[width=4.2in]{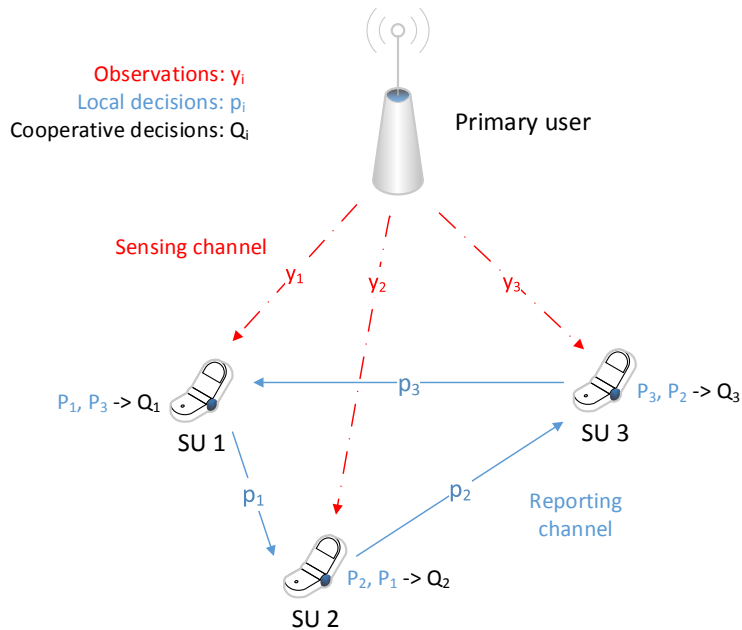}
\caption{Application scenario~\uppercase\expandafter{\romannumeral2}: cooperative sensing} \label{app2}
\end{figure}

In order to provide gigabit transmission rate, future wireless networks must use a large amount of spectrum resources. However, the scarcity of the radio spectrum coupled with the existing, fixed spectrum assignment policies, has motivated the need for new, dynamic spectrum access mechanisms to efficiently exploit the spectral resources. Cognitive radio~(CR) is one highly promising technique to achieve such dynamic exploitation of the radio spectrum. In CR networks, unlicensed, secondary users~(SUs), can sense the environment and change their parameters to access the spectrum of licensed, primary users~(PUs), while maintaining the interference to the PUs below a tolerable threshold. Such spectrum sensing is an integral part of any CR network. Indeed, reaping the benefit of CR is contingent upon deploying smart and efficient spectrum sensing policies. Here, we consider the cooperative sensing scheme, in which nearby SUs exchange their local sensing results and then make collaborative decisions on the detection of PUs. There are three categories of cooperative spectrum sensing, based on how cooperating SUs share their sensing data
in the network: centralized, distributed, and relay-assisted~\cite{ALB-2011}. Here, we consider the distributed case, where SUs communicate among themselves with no fusion center or relay, as seen in Fig.~\ref{app2}. Traditionally, the distributed approach adopts non-cooperative schemes for its simplicity~\cite{LYH-2009}. In this paper, we will show how OCF games can be used to introduce cooperation between SUs, and thus, improve the overall sensing performance~\cite{WSHS-2013}.

Consider a cognitive radio network with multiple SUs equipped with energy detectors and a single PU far away from them. In this network, the SUs can individually and locally decide on the presence or absence of the PU via their own detectors. Then, they can cooperate with one another by exchanging their local decisions via a reporting channel. At last, each SU combines its local decision with the received decisions and decides whether or not the PU is present. Note that the SUs may have different local detectors with different detection threshold, their missed detection probabilities and false alarm probabilities may be different.

\begin{figure}[!t]
\centering
\includegraphics[width=4.0in]{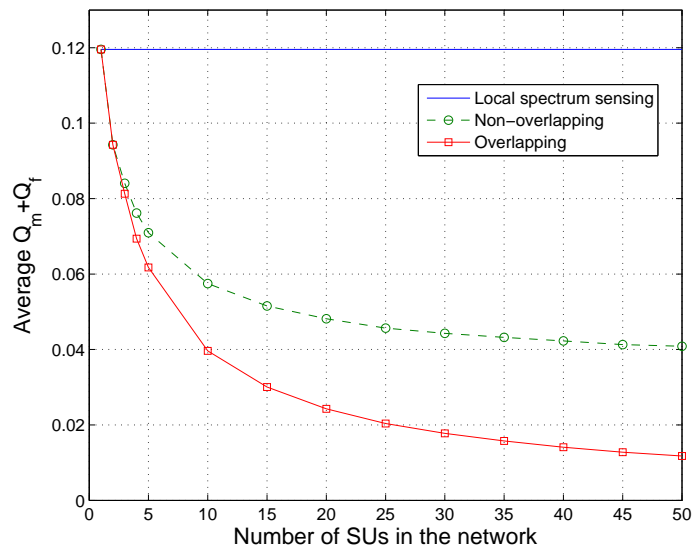}
\caption{Network sensing performance as a function of the number of SUs.} \label{cr}
\end{figure}

In process of cooperative sensing, each SU in the system needs to collect local decisions from other SUs, and thus, each SU represents a sensing task to be accomplished via the cooperation of SUs. However, the bandwidth of the report channel is limited for every SU, which is usually not enough to transmit to all other SUs, especially for those SUs with bad channel conditions. Thus, a SU needs to decide how to cooperate with other SUs by efficiently allocating its limited bandwidth to the transmissions of different SUs. Therefore, this is $K$-task OCF game, where each task is presented by a coalition composed of a head SU that collects local decisions and other SUs that report to it. The value of the coalition is then given by the sensing performance of the head SU, which can be calculated based on the adopted fusion rule. We utilize the developed algorithm in Table~\ref{K-task} and show its performance in Fig.~\ref{cr}.

In Fig.~\ref{cr}, we compare the developed algorithm with the local sensing method and the non-overlapping method. In the local sensing method, the SUs use their local sensing decision without any information exchange. In the non-overlapping method, the SUs form non-overlapping coalitions and each SU only share information inside the coalition it joins. Clearly, the developed algorithm outperforms both methods. The incorrect probability decreases to $1/12$ of the local sensing method and $1/4$ of the non-overlapping method. Also, the improvement becomes larger as the number of SUs increases.

\section{Other Applications}%

OCF game are quite suitable for modeling the future wireless networks, in which the wireless nodes are dense, self-organizing, and cooperative. In this section, we briefly discuss other potential applications of OCF games and then summarize the applications in Table~\ref{app}.

\begin{table}
\renewcommand{\arraystretch}{1.0}
\caption{Applications of OCF games} \label{app} \centering
\begin{tabular}{|m{22mm}|m{15mm}|m{15mm}|m{35mm}|m{40mm}|m{20mm}|}

\hline \textbf{Application} & \textbf{Player} & \textbf{Resources} & \textbf{Coalition} & \textbf{Coalition Value} & \textbf{Type} \\
\hline Radio resource allocation & Small cells & Available RBs & Radio coordination among the RBs from different small cells & Total throughput of the coordinated RBs considering all the potential interference & $K$-coalition \\
\hline Cooperative spectrum sensing & SUs & Signaling bits & A specific SU and the SUs that report to this SU & the cooperative sensing performance of this specific secondary user & $K$-task \\
\hline Multi-radio traffic offloading & Multi-mode devices & User traffic & A specific base station and the traffic contributed by different devices & A function reflecting the user experience & $K$-task \\
\hline CoMP & Base stations & Radio resources & A cell-edge user and the resources contributed by different base stations & The throughput of the cell-edge user & $K$-task \\
\hline Virtual MIMO & Mobile phones & Radio resources & Cooperative users forming a virtual MIMO group & the MIMO link rate & $K$-coalition \\
\hline Smartphone sensing & Smartphones & Battery energy & A task and the energy devoted by different smartphones & The task utility & $K$-task \\

\hline
\end{tabular}
\end{table}

\subsection{Multi-Radio Traffic Offloading}

Cellular networks are constantly evolving into their next generation. However, the former systems are not entirely replaced by the new systems. In fact, it is expected that different networks will coexist for a long time, and, thus, mobile phones will be multi-mode terminals that enable communications over different radio access technologies (RATs). In order to fully explore their network investments, the operators must intelligently offload their network traffic over different RATs. Developing such offloading schemes, which must consider the demands and access authorities of different users, the transmitting rates of different technologies, and the deployment and load of different base stations, is quite challenging for a large number of users and base stations. However, one can use the proposed $K$-task OCF game to model this problem.

In the OCF game model, the mobile users can distribute their traffic into different base stations in different networks. A coalition here represents a base station as well as the traffic devoted from different mobile users. The coalition value can be simply defined as the total throughput of this base station with channel and technology limitations, or a sophisticated function reflecting the user experience, which considers the delay and rate experienced by the users, and the cost and energy efficiency of the network. Using the developed algorithm in Table~\ref{K-task}, the user traffic can be intelligently distributed among different networks with high network performance in the sense of the defined value function.

\subsection{Cooperative Communications}

In order to increase the performance of cell-edge users, coordinated multipoint (CoMP) transmission has been proposed, in which the signals of multiple base stations are coordinated to serve a cell-edge user. Since there are multiple cell-edge users, the base stations should allocate their radio resources among these users. It is a challenging optimization problem, since the channel conditions, traffic demands and radio resources are different for different users and base stations. However, we can model this problem using a $K$-task OCF game.

In the OCF game model, the base stations can freely allocate their radio resources to different users, including bandwidth, power and antenna resources. A coalition represents a cell-edge user as well as the radio resources devoted from different base stations. The coalition value is defined as the throughput of this cell-edge user. Thus, using the developed algorithm in Table~\ref{K-task}, the radio resources of base stations can be efficiently distributed among different cell-edge users.

Another related application is the cooperation between user devices. In order to increase their transmission rate, nearby users may group together to use virtual MIMO transmissions. The MIMO link rate is generally increasing with the number of cooperated users, while the marginal increase is decreasing due to the increasing distance between different users. Thus, a user may want to allocate its radio resources among different cooperative groups, so as to maximize its individual throughput. This problem can be modeled via a $K$-coalition OCF game, in which a coalition represents a virtual MIMO group and the coalition value is the MIMO link rate. Using the developed algorithm in Table~\ref{K-coalition}, the radio resources of users can be efficiently distributed among different virtual MIMO groups.

\subsection{Smartphone Sensing}

In recent years, smartphones are equipped with more and more sensors. These powerful sensors allow public departments or commercial companies to accomplish large-area sensing tasks via individual smartphones. These tasks often require collecting data in a large area, and thus, a huge number of smartphones may be involved. Based on the task itself and the geographic locations of smartphones, different tasks may require different amount of energy and provide different payoffs for different smartphones. A smartphone user must decide to which tasks he should devote the limited energy. Therefore, we can model this problem with the studied $K$-task OCF game, in which each coalition represents a task and the energy devoted from different smartphones, and the coalition value is given by the task utility. Using the developed algorithm in Table~\ref{K-task}, the smartphone users can efficiently allocate their energy into different sensing tasks.

\section{Conclusions}%

In this paper, we have introduced the framework of overlapping coalition formation games as a tool to model and analyze the communication scenarios in future networks. In particular, we have defined two subclasses, namely $K$-coalition and $K$-task OCF games, and we have developed polynomial algorithms to achieve an o-stable outcome. Subsequently, we have presented, in detail, how OCF games can be used to address challenging problems in two application domains: radio resource allocation in HetNets and cooperative sensing. In addition, we have discussed some other potential applications of OCF-games, including multi-radio traffic offloading, cooperative communications, and smartphone sensing. Finally, we envision that the use of the OCF game framework will play an important role in 5G networks, particularly, as the network becomes more dense, decentralized and self-organizing.


\begin{thebibliography}{50}

\bibitem{Cisco-2014}
Cisco, ``Visual Networking Index," white paper at Cisco.com, Feb.~2014.

\bibitem{ABCHLSZ-2014}
J.~G.~Andrews, S.~Buzzi, W.~Choi, S.~Hanly, A.~Lozano, A.~C.~K.~Soong, and J.~C.~Zhang, ``What Will 5G Be?" \emph{IEEE Journal on Selected Areas in Communications}, vol.~32, no.~6, pp.~1065~1081, Jun.~2014.

\bibitem{A-2013}
J.~G.~Andrews, ``Seven Ways that HetNets Are a Cellular Paradigm Shift," \emph{IEEE Communications Magzine}, vol.~51, no.~3, pp.~136-144, Mar.~2013

\bibitem{OHI-2012}
O.~Onireti, F.~Heliot, and M.~A.~Imran, ``On the Energy Efficiency-Spectral Efficiency Trade-Off in the Uplink of CoMP System," \emph{IEEE Transactions on Wireless Communications}, vol.~11, no.~2, pp.~556-561, Feb.~2012.

\bibitem{PBSDL-2013}
F.~Pantisano, M.~Bennis, W.~Saad, M.~Debbah, and M.~Latva-aho, ``Interference Alignment for Cooperative Femtocell Networks: A Game-Theoretic Approach," IEEE Transactions on Mobile Computing, vol.~12, no.~11, pp.~2233-2246, Nov.~2013.

\bibitem{KLNK-2014}
Y.~Kawamoto, J.~Liu, H.~Nishiyama, and N.~Kato, ``An Efficient Traffic Detouring Method by Using Device-to-Device Communication Technologies in Heterogeneous Network.", \emph{IEEE Wireless Communications and Networking Conference}, Istanbul, Turkey, Apr.~2014.

\bibitem{FWGLZ-2013}
W.~Feng, Y.~Wang, N.~Ge, J.~Lu, and J.~Zhang, ``Virtual MIMO in Multi-Cell Distributed Antenna Systems: Coordinated Transmissions with Large-Scale CSIT," \emph{IEEE Journal on Selected Areas in Communications}, vol.~31, no.~10, pp.~2067-2081, Oct.~2013.

\bibitem{HNSBH-2011}
Z.~Han, D.~Niyato, W.~Saad, T.~Basar, and A.~Hjorungnes, \emph{Game Theory in Wireless and Communication Networks: Theory, Models and Applications}, Cambridge University Press, UK, 2011.

\bibitem{SHDHB-2009}
W.~ Saad, Z.~Han, M.~Debbah, A.~Hjorungnes, and T.~Basar, ``Coalitional Game Theory for Communication Networks," \emph{IEEE Signal Processing Magazine, Special Issue on Game Theory}, vol.~26, no.~5, pp.~77-97, Sep.~2009.

\bibitem{SHDH-2008}
W.~Saad, Z.~Han, M.~Debbah, and A.~Hjorungnes, ``A Distributed Merge and Split Algorithm for Fair Cooperation in Wireless Networks," in \emph{Proceedings of International Conferernce on Communications, Workshop Cooperative Communications and Networking}, Beijing, China, May~2008.

\bibitem{SHDHB-2009b}
W.~Saad, Z.~Han, M.~Debbah, A.~Hjorungnes, and T.~Basar, ``Coalitional Games for Distributed Collaborative Spectrum Sensing in Cognitive Radio Networks," in \emph{Proceedings of IEEE INFOCOM}, Rio de Janeiro, Brazil, Apr.~2009.

\bibitem{SHBDH-2009}
W.~Saad, Z.~Han, T.~Basar, M.~Debbah, and A.~Hjorungnes, ``Physical Layer Security: Coalitional Games for Distributed Cooperation," in \emph{Proceedings of 7th International Symposium of Modeling and Optimization in Mobile, Ad Hoc, and Wireless Networks}, Seoul, South Korea, Jun.~2009.

\bibitem{HP-2009}
Z.~Han, H.~V.~Poor, ``Coalition Games with Cooperative Transmission: A Cure for the Curse of Boundary Nodes in Selfish Packet-Forwarding Wireless Networks," \emph{IEEE Transactions on Communications}, vol.~57, no.~1, pp.~203-213, Jan.~2009.

\bibitem{SHBDH-2010}
W.~Saad, Z.~Han, T.~Basar, M.~Debbah, A.~Hjorungnes, ``Hedonic Coalition Formation for Distributed Task Allocation among Wireless Agents," \emph{IEEE Transactions on Mobile Computing}, vol.~10, no.~9, pp.~1327-1344, Dec.~2010.

\bibitem{SHHNH-2011}
W.~Saad, Z.~Han, A.~Hjorungnes, D.~Niyato, and E.~Hossain, ``Coalition Formation Games for Distributed Cooperation Among Roadside Units in Vehicular Networks," \emph{IEEE Journal on Selected Areas in Communications}, vol.~29, no.~1, pp.~48-60, Jan.~2011.

\bibitem{PBSVL-2011}
F.~Pantisano, M.~Bennis, W.~Saad W, R.~Verdone, M.~Latva-aho, ``Coalition Formation Games for Femtocell Interference Management: A Recursive Core Approach," in \emph{Proceedings of Wireless Communications and Networking Conference}, Quintana-Roo, Mexico, Mar.~2011.

\bibitem{WCZW-2012}
D.~Wu, Y.~Cai, L.~Zhou, and J.~Wang, ``A Cooperative Communication Scheme Based on Coalition Formation Game in Clustered Wireless Sensor Networks," \emph{IEEE Transactions on Wireless Communications}, vol.~11, no.~3, pp.~1190-1200, Feb.~2012.

\bibitem{WSH-2013}
T.~Wang, L.~Song and Z.Han, ``Coalitional Graph Games for Popular Content Distribution in Cognitive Radio VANETs," \emph{IEEE Transactions on Vehicular Technology}, vol.~62, no.~8, pp.~4010-4019, Jun.~2013.

\bibitem{ZSHS-2014}
Z.~Zhang, L.~Song, Z.~Han, and W.~Saad, ``Coalitional Games with Overlapping Coalitions for Interference Management in Small Cell Networks," \emph{IEEE Transactions on Wireless Communications}, vol.~13, no.~5, pp.~2659-2669, May~2014.

\bibitem{WSHS-2013}
T.~Wang, L.~Song, Z.~Han, and W.~Saad, ``Overlapping Coalitional Games for Collaborative Sensing in Cognitive Radio Networks," in \emph{Proceedings of Wireless Communications and Networking Conference}, Shanghai, China, Apr.~2013.

\bibitem{DWSH-2013}
B.~Di, T.~Wang, L.~Song, and Z.~Han, ``Incentive Mechanism for Collaborative Smartphone Sensing using Overlapping Coalition Formation Games," in \emph{Proceedings of Global Communications Conference}, Atlanta, GA, Dec.~2013.

\bibitem{BSSVDC-2013}
M.~Bennis, M.~Simsek, W.~Saad, S.~Valentin, M.~Debbah, and A.~Czylwik, ``When Cellular Meets WiFi in Wireless Small Cell Networks," \emph{IEEE Communications Magazine, Special Issue on Heterogeneous Networks}, vol.~51, no.~6, Jun.~2013.

\bibitem{ZANM-2013}
T.~Zahir, K.~Arshad, A.~Nakata, and K.~Moessner, ``Interference Management in Femtocells," \emph{IEEE Communications Surveys and Tutorials}, vol.~15, no.~1, pp.~293-311, Feb.~2013.

\bibitem{Andrews-2005}
J.~G.~Andrews, ``Interference Cancellation for Cellular Systems: A Contemporary Overview," \emph{IEEE Wireless Communications}, vol.~2, no.~3, pp.19-29, Apr.~2005.

\bibitem{SS-2011}
P.~Sangkyu and B.~Saewoong, ``Dynamic Inter-cell Interference Avoidance in Self-organizing Femtocell Networks," in \emph{Proceedings of IEEE International Conference on Communications}, Tokyo, Japan, Jun.~2011.

\bibitem{HT-2010}
W.~Hong and Z.~Tsai, ``On the Femtocell-based MVNO Model: A Game Theoretic Approach for Optimal Power Setting," in \emph{Proceedings of IEEE 71st Vehicular Technology Conference}, Taipei, Taiwan, May~2010.

\bibitem{ALB-2011}
I.~F.~Akyildiz, B.~F.~Lo, R.~Balakrishnan, ``Cooperative Spectrum Sensing in Cognitive Radio Networks: A Survey." \emph{Physical Communication}, vol.~4, no.~1, pp.~40-62, Mar.~2011.

\bibitem{LYH-2009}
Z.~Li, F.~Yu, and M.~Huang, ``A Cooperative Spectrum Sensing Consensus Scheme in Cognitive Radios," in \emph{Proceedings of IEEE Infocom}, Rio de Janeiro, Brazil, Apr.~2009.

\bibitem{CEMJ-2010}
G.~Chalkiadakis, E.~Elkind, E.~Markakis, and N.~R.~Jennings, ``Cooperative Games with Overlapping Coalitions," \emph{Journal of Artificial Intelligence Research 39}, vol.~39, no.~1, pp.~179-216, Sep.~2010.

\bibitem{ZE-2011}
Y.~Zick and E.~Elkind, ``Arbitrators in Overlapping Coalition Formation Games," in \emph{Proceedings of 10th International Conference on Autonomous Agents and Multiagent Systems}, Taipei, Taiwan, May~2011.

\bibitem{ZCE-2012}
Y.~Zick, G.~Chalkiadakis, and E.~Elkind, ``Overlapping Coalition Formation Games Charting the Tractability Frontier," in \emph{Proceedings of 10th International Conference on Autonomous Agents and Multiagent Systems}, Valencia, Spain, Jun.~2012.

\bibitem{CGYZ-2014}
X.~Chen, X.~Gong, L.~Yang, and J.~Zhang, ``A Social Group Utility Maximization Framework with Applications in Database Assisted Spectrum Access," in \emph{Proceedings of IEEE INFOCOM}, Toronto, Canada, Apr.~2014.

\end{thebibliography}
\end{document}